\documentclass[aps,prl,showpacs]{revtex4}
\usepackage{graphicx}
\usepackage{verbatim}

\newcommand{\req}[1]{(\ref{#1})}
\newcommand{\be}{\begin{equation}}
\newcommand{\ee}{\end{equation}}
\newcommand{\bea}{\begin{eqnarray}}
\newcommand{\eea}{\end{eqnarray}}
\newcommand{\dd}{\textrm{d}}

\newcommand{\cro}[1]{\left[#1\right]}

\newcommand{\BE}{\begin{eqnarray}}
\newcommand{\EE}{\end{eqnarray}}
\newcommand{\BEn}{\begin{eqnarray*}}
\newcommand{\EEn}{\end{eqnarray*}}
\newcommand{\barr}{\begin{array}}
\newcommand{\earr}{\end{array}}

\newcommand{\bit}{\begin{itemize}}      
\newcommand{\eit}{\end{itemize}}
\newcommand{\bc}{\begin{center}}
\newcommand{\ec}{\end{center}}
\newcommand{\ben}{\begin{enumerate}}    
\newcommand{\een}{\end{enumerate}}

\newcommand{\eps}{\epsilon}

\newcommand{\erf}{{\rm erf}}

\begin{document}

\title{News and price returns from threshold behaviour and vice-versa: exact solution of an agent-based market model}
\author{Damien Challet}
\affiliation{Nomura Centre for Quantitative Finance, Mathematical Institute, Oxford University, 24--29 St Giles', Oxford OX1 3LB, United Kingdom}
\email{challet@maths.ox.ac.uk}
\date{\today}


\begin{abstract}
Starting from an exact relationship between news, threshold and price return distributions in the stationary state, I discuss the ability of the Ghoulmie-Cont-Nadal model of traders to produce fat-tailed price returns. Under normal conditions, this model is not able to transform Gaussian news into fat-tailed price returns. When the variance of the news so small that only the players with zero threshold can possibly react to news, this model produces Levy-distributed price returns with a $-1$  exponent. In the special case of  super-linear price impact functions, fat-tailed returns are obtained from well-behaved news. 
\end{abstract}

\maketitle

Few agent-based models of financial markets are amenable to mathematical analysis. Notable exceptions include percolation-based imitations models \cite{ContBouchaud}, behavioural switching models \cite{Alfarano,Hommes_HAM_review}, and Minority Game-based models \cite{MGbook,CoolenBook}. Mathematical solutions are valuable tools for exploring the mechanisms underlying financial market dynamics, particularly in a view to understand the causes of unpredictability and large fluctuations. Thresholds mechanisms, whereby the agents react to events exceeding their personal tolerance levels, provide an indirect way to avoid forcing the agents to trade at each time step, and also implicitly define timescales. 

A new simple Markovian model of agents that adjust their thresholds on past volatilities was recently introduced and exhibited promising features \cite{Cont2}. Here, I derive exact relationships between price return, thresholds and news distribution in the limit of infinite number of agents, and discuss in details when this model produces fat-tailed returns from well-behaved news distributions. 

The model works as follows \cite{Cont2}: there are $N$ agents with individual threshold $\theta_i$, $i=1,\cdots,N$, who monitor a single source of external news. At time $t$ the news intensity and sign is $\eps(t)$, drawn from the symmetric distribution $\rho(\eps)$; a positive $\eps$ corresponds to good news, and reversely. Only the agents whose thresholds are smaller than the absolute value of the news react to it by placing a order to buy if $\eps(t)>0$ and reversely. Mathematically, if $\phi_i(t)$ denotes the action of agent $i$ at time $t$,
\be
\phi_i(t)=\Theta(\eps(t))\Theta(\eps_t-\theta_i(t))-\Theta(-\eps(t))\Theta(\theta_i-\eps(t)),
\ee
where $\Theta(x)$ is the Heaviside function; note that $\phi_i(t)\in\{-1,0,1\}$. This gives rise to global demand imbalance $\Phi(t)=\sum_i\phi_i(t)$. The price evolves according to
\be
\log p(t+1)=\log p(t) + r(t),
\ee where $r$ is called the price return. Its relationship with demand imbalance is assumed to be fully characterised by a constant price impact function $g$
\be\label{return}
r(t)=g(\Phi(t)/N).
\ee
 At each time step, agent $i$ may re-adjust his threshold $\theta_i$ with probability $s$ to $|r(t)|$. In that way, the agents are given minimal learning capabilities, tracking changes of volatility with a reaction time scale of order $1/s$. It is worth noting that the agents have no influence on the news process. The inventors of this model produced evidence that the agents transform Gaussian news into price return distributions $P(r)$ whose shape and properties depend on the parameters $N$, $s$ and $\rho$, some of them seemingly not Gaussian anymore  \cite{Cont2}.

\section*{Stationary state}

Since $\rho(\eps)$ is symmetric, the model can be simplified by assuming that replacing $\eps$ by $\eps'=|\eps|$ and and $\rho(\eps')$ by $\rho'(\eps')=2\rho(|\eps|)$. Dropping the primes, the dynamics of individual actions reduces to
\be
\phi_i(t)=\Theta(\eps_t-\theta_i(t)).
\ee
Because of Eq. \req{return}, since $\Phi$ is discrete for finite $N$, $r(t)$ and $\theta_i$ take discrete values.
The master equation for $P(\theta,t)$ is therefore
\be\label{master}
P(\theta,t+1)=(1-s)P(\theta,t)+s\delta[\theta-g(\sum_{\theta'\le\eps_t}P(\theta',t))]
\ee
Letting $\tau=\theta/N$ and $\eps'=\eps/N$, and dropping the prime, one can transform this problem into a finite continuous interval ($r$,  $\tau$ and $\eps$ belong to $[0,1]$), resulting in
\be
P(\tau,t+1)=(1-s)P(\tau,t)+s \delta[\tau-g(\int_0^{\eps_t} \dd \tau' P(\tau',t))].
\ee
By definition, $P(\tau,t)$ becomes $P(\tau)$ in the stationary state; the final step is to average over all possible $\eps$, which yields 
\be\label{P(tau)}
P(\tau)=\int\dd\eps \rho(\eps) \delta(\tau-g[F(\eps)])=\frac{\rho[F^{-1}(g^{-1}(\tau))]}{P[F^{-1}(g^{-1}(\tau))])g'(g^{-1}(\tau))}.
\ee
It should be noted that the disappearance of $s$ in this equation is a sign that $s$ is a time scale; the auto-correlation function of $r(t)$ stays positive for a time scale of order $1/s$, as can be seen in the original paper. In the stationary state, since $r=g(F(\eps))$, the pdf of $r$ has exactly the same functional form, because the thresholds are adjusted to the last price return with probability $s$. The multiple appearances of $P(\tau)$ under various forms in the right hand side of Eq \req{P(tau)} is due to the self-referential nature of the model.

The dynamics produces $P(\tau)$ from $\rho(\eps)$. However, Eq \req{P(tau)} seems quite difficult to solve; it is much easier to find which $\rho(\eps)$ corresponds to a given $P(\tau)$, since 
\be
\rho(\eps)= P[g(F(\eps))]g'(F(\eps))P(\eps)
\ee
does not depend on itself in an intricate way.

\section*{Linear price impact function}

The original model reports simulations with $g(r)=\lambda r$, which simplifies the discussion. Without loss of generality, one takes $\lambda=1$. The original model also assumes Gaussian news arrival, that is, $\rho(\eps)=2{\cal N}(0,D)$, $\eps\ge0$.  While it would hard to find $P(\tau)$ that leads exactly to such a distribution, $P(\tau)=2{\cal N}(0,D)$ corresponds to
\be
\rho(\eps)=\frac{2}{D^2\pi}e^{-\frac{\erf\cro{\frac{\eps}{\sqrt{2}D}}^2}{2D^
2}}e^{-\frac{\eps^2}{2D^2}}
\ee
In particular, for $\eps\ge D$, $\rho(\eps)$ approximates very well a Gaussian. It is therefore possible to conclude that Gaussian news produces Gaussian-tailed price return distributions. Numerical simulations show that Gaussian news can produce one peak or two symmetric peaks with or without one central Levy distributed peak (in finite size systems) whose origin is explained below. 

\begin{figure}
\centerline{\includegraphics*[width=0.4\textwidth]{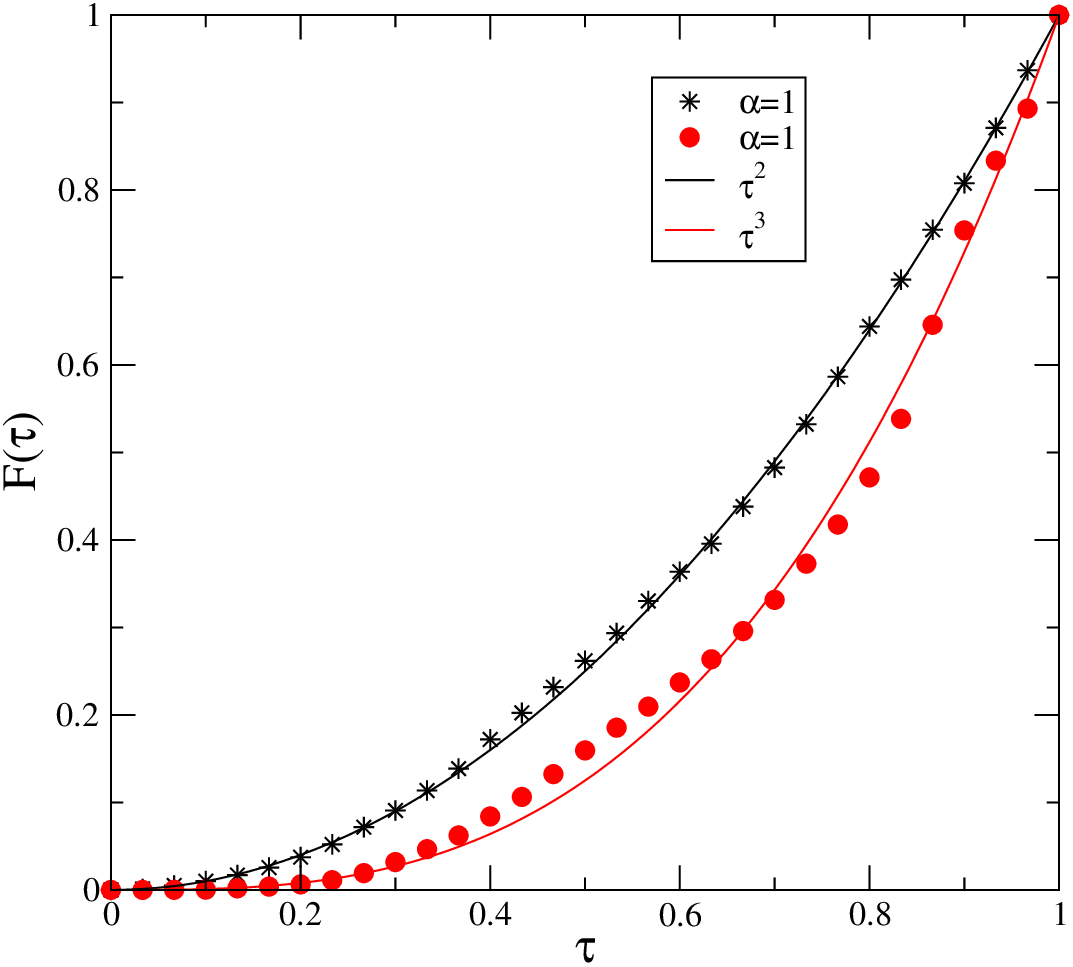}\includegraphics*[width=0.4\textwidth]{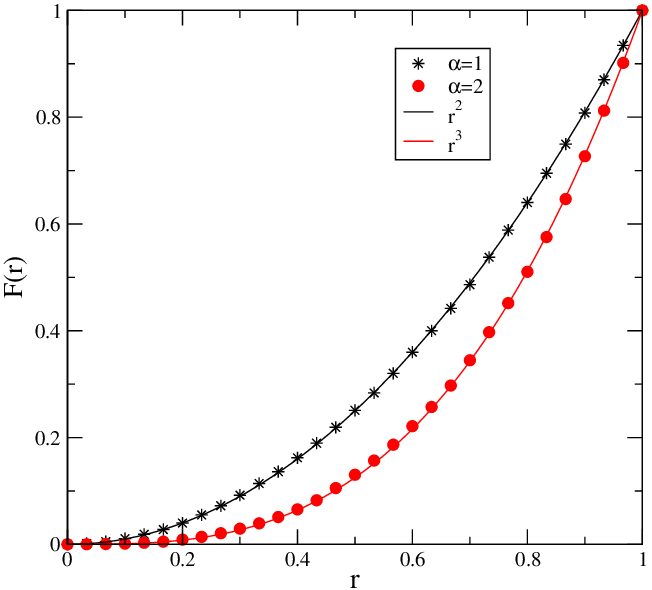}}
\caption{Threshold (left) and price return (right) cumulative distribution function $F(\tau)$, respectively $F(r)$ obtained with $\rho(\eps)=(\alpha+1)^2\eps^{\alpha(\alpha+2)}$ for $\alpha=1$ (black symbol and line) and $\alpha=2$ (red symbol and line). $N=10000$, $s=0.001$, $100000$ iterations; $F(\tau)$ is averaged over 9 equally spaced snapshots in the last half of the iterations.}
\label{Ftau_rho-pw}
\end{figure}

Nevertheless, one might question the assumption that news are Gaussian. For instance damages caused by natural disasters have fat tails \cite{EmbrechtsExtremalEvents}. The power-law case yields a more direct relationship between news and price returns: $P(r)=(\alpha+1)\tau^{\alpha}$ ($\alpha>-1$) corresponds to  $\rho(\eps)=(\alpha+1)^2\eps^{\alpha(\alpha+2)}=(\alpha+1)\eps^\beta$ for $r$ and $\eps\,\in[0,1]$: it is therefore impossible to produce $\alpha<0$ from $\beta>0$ with linear price impact functions, hence, well-behaved news cannot trigger fat-tailed price returns; however, fat-tailed news do trigger fat-tailed prices. Fig \ref{Ftau_rho-pw} confirms the validity of the above calculations: numerical and theoretical values of $P(r)$ match perfectly. The case of $P(\tau)$ is interesting: even after $10^5$ time steps, there is a discrepancy between numerical simulations and theoretical predictions; this comes from the temporal correlations of the thresholds over time scales of order $1/s$, which causes the presence of peaks in any snapshot of $P(\tau)$. When one averages $P(\tau)$ over several snapshots, one finds that the theoretical prediction and the numerical simulations match better. Alternatively, one could plot the distribution of the values taken by the threshold of a given agent. As a final example, an exponential distribution of thresholds $P(\tau)=e^{-\tau}$, $\tau\ge0$, corresponds to $\rho(\eps)= e^{ e^{-\eps}}e^{-\eps}$.

So far, the discussion has only addressed the model in the limit of infinite systems. However, the behaviour of the model changes completely for finite size if $D\ll1/N$ and $sN>1$. In that case, it is easy to convince oneself that the only situation yielding non-zero price returns occurs when $P(0)\ne0$; in other words, in the case of very small news, the only agents who react to them are infinitely susceptible, as they react in fact to any news amplitude. The mechanism is the following: the smallest non-zero threshold is $1/N>\eps_t$ for all $t$. Assume that $P(\theta=0,t_0)=0$, which implies that $r(t_0)=0$. Then on average $sN$ agents adjust their threshold to $\theta=0$. At the next time step, all of them react to any news amplitude, which leads to an average price return $r(t_0+1)=s$. Because of the threshold adjustment process, the number of agents with a zero threshold follows a multiplicative random walk $r(t+1)=r(t)\eta(t)$ where ${\cal P}(\eta)\simeq{\cal N}(1-s,s(1-s)/N)$, with a barrier at $1/N$; when $P(0,t')<1/N$, $r$ jumps back on average to $s$. Numerical simulations fully confirm this mechanism. They also show that ${ P}(r)\propto r^{-1}$ (and of course $P(\tau)\propto \tau^{-1}$), with an upper cut-off at about $s$. Multiplicative random walks with lower boundary are well known to give rise to power-law distributed density functions whose exponent $\gamma$ is given by (see for instance \cite{MZMPortfolio})
\be
\int\dd \eta P(\eta)\eta^{\gamma-1}=1.
\ee
In the large $N$ limit, ${\cal P}(\eta)$ converges to a Dirac function, which leads to $(1-s)^{\gamma-1}=$1, hence $\gamma=1$. The normalisation of $P(r)$ is not problematic, since $r$ is bounded from above. Intermediate cases occur when $sN\sim 1$: the centre of the distribution is of Levy type with $-1$ exponent, and is truncated.

\begin{figure}
\centerline{\includegraphics*[width=0.35\textwidth]{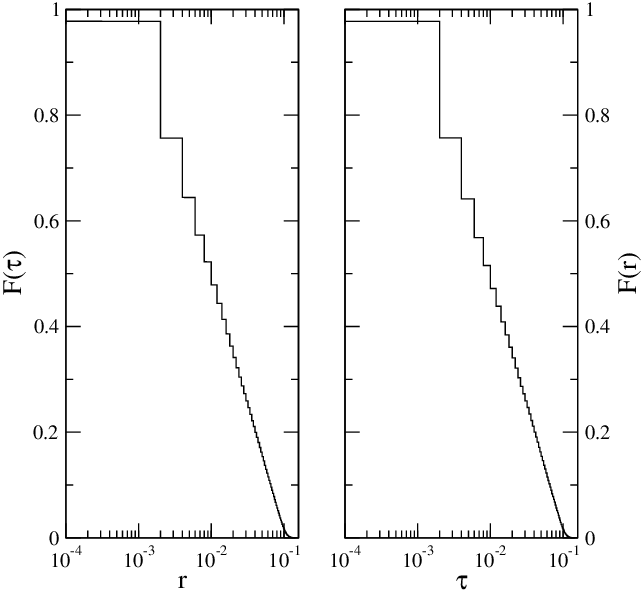}}
\label{Prtau}
\caption{Cumulative distribution of $r$ and $\tau$ for $N=5000$, $D=0.1/N$, $s=0.1$,  $10^5$ iterations.}
\end{figure}

\section*{Non-linear price impact functions}

Average real market price impact functions are reportedly concave functions, either $g(x)=x^\gamma$ with $\gamma<1$ \cite{StanleyImpact,FarmerImpact} or $g(x)=\log x$ \cite{BouchaudLimit3}. It will be enough to discuss the power-law case. Using $P(r)=(\alpha+1)\tau^{\alpha}$ ($\alpha>-1$), one finds that 
\be
\rho(\eps)=\gamma(\alpha+1)^2\eps^{\gamma(\alpha+1)^2-1}=\gamma(\alpha+1)^2\eps^\beta.
\ee
Equivalently, $\alpha=\sqrt{\frac{\beta+1}{\gamma}}-1$. The condition for power-law distributed price returns ($\alpha<0$) is $\beta<\gamma-1$. Therefore, sublinear impact functions dampen $\rho$, and fat-tailed returns are only obtained from sufficiently fat-tailed news; indeed, if $\gamma-1<\beta<0$, the news are fat-tailed, but the price returns are not. A superlinear $g$ on the other hand makes it possible to have a well-behaved $\rho$ and fat-tailed price returns, but, as stated above, average price impact functions are not convex functions.

\section*{Conclusion}

In short, in the $N\to\infty$ limit, the above derivation not only shows the crucial importance of the functional form of the news on the price properties, but also that this model produces very fat-tailed price returns in the limit of infinitesimally small news with exponent $1$, and only for finite size systems; the other possibility for fat-tailed price returns arises when the price impact function is super-linear. Therefore, under realistic assumptions, this model cannot transform Gaussian news into fat-tailed price returns.

I wish to thank Andreas Ipsen for stimulating discussions and an anonymous referee for useful suggestions.



\bibliography{biblio}

\end{document}